\documentstyle[aps,prb,graphicx,twocolumn]{revtex}

\begin{document}
\draft
\title{Quantum-dot lithium in zero magnetic field: Electronic properties, thermodynamics, and a liquid-solid transition in the ground state
}

\author{S. A. Mikhailov\footnote{Electronic address: sergey.mikhailov@physik.uni-augsburg.de}}
\address{Theoretical Physics II, Institute for Physics, University of Augsburg, 86135 Augsburg, Germany}

\date{\today}
\maketitle

\begin{abstract}
Energy spectra, electron densities, pair correlation functions and heat capacity of a quantum-dot lithium in zero external magnetic field (a system of three interacting two-dimensional electrons in a parabolic confinement potential) are studied using the exact diagonalization approach. A particular attention is given to a Fermi-liquid -- Wigner-solid transition in the ground state of the dot, induced by the intra-dot Coulomb interaction.  
\end{abstract}


\section{Introduction}
\label{sec:intro}

Quantum dots \cite{Jacak97} are artificial electron systems (ES) realizable in modern semiconductor structures. In these systems two-dimensional (2D) electrons move in the plane $z=0$ in a lateral confinement potential $V(x,y)$. The typical length scale $l_0$ of the lateral confinement is usually larger than or comparable with the effective Bohr radius $a_B$ of the host semiconductor. The relative strength of the electron-electron and electron-confinement interaction, given by the ratio $\lambda\equiv l_0/a_B$, can be varied, even experimentally, in a wide range, so that the dots are used to be treated as artificial atoms with tunable physical properties. Experimentally, quantum dots were intensively studied in recent years, using a variety of different techniques, including capacitance \cite{Ashoori96}, transport \cite{Kouvenhouven97}, far-infrared \cite{Heitmann93} and Raman spectroscopy \cite{Schuller98}. 

From the theoretical point of view, quantum dots are ideal physical objects for studying effects of electron-electron correlations. Different theoretical approaches, including analytical calculations \cite{Taut93,Taut94,Turbiner94,Dineykhan97}, exact diagonalization \cite{Maksym90,Merkt91,Pfannkuche91,Wagner92,Hawrylak93,Hawrylak93b,MacDonald93,Yang93,Pfannkuche93,Palacios94,Peeters96,Ezaki97,Eto97,Creffield99,Reimann00,Yannouleas00}, quantum Monte Carlo (QMC) \cite{Maksym96,Bolton96,Harju98,Harju99a,Harju99b,Egger99a,Egger99b,Pederiva00,Filinov01}, density functional theory \cite{Hirose99,Steffens98a,Steffens98b,Steffens98c,Wensauer00} and other methods \cite{Maksym95,Ruan95,Hausler96,Yannouleas99,Hausler00,Taut00,Reusch01}, were applied to study their properties, for a recent review see Ref. \cite{Maksym00}. Until recently most theoretical work was performed in the regime of strong magnetic fields, when all electron spins are fully polarized. In the past three years a growing interest is observed in studying the quantum dot properties in zero magnetic field $B=0$ \cite{Creffield99,Reimann00,Egger99a,Egger99b,Filinov01,Hausler00,Reusch01}. The aim of these studies is to investigate the Fermi liquid -- Wigner solid crossover in the dots, at a varying strength of Coulomb interaction. Detailed knowledge of the physics of such a crossover in microscopic dots could be compared with that obtained for macroscopic 2DES \cite{Tanatar89,Chui95} and might shed light on the nature of the metal-insulator transition in two dimensions \cite{Abrahams01}. 

So far, full energy spectra of an $N$-electron parabolic quantum dot in zero magnetic field, as a function of the interaction parameter $\lambda$, were published only for $N=2$ (quantum-dot helium \cite{Merkt91}). For larger $N$ a number of results for the ground state energy of the dots were reported at separate points of the $\lambda$ axis. Sometimes, however, results obtained by different methods contradict to each other, and full understanding of physical properties of $N$-electron dots at $B=0$ has not yet been achieved. 

Even for three electrons in a parabolic confining potential (quantum-dot lithium) available in the literature results are somewhat confusing and do not give a clear and exact picture of the ground state of the dot. Egger et al. reported in \cite{Egger99a} on a transition from a partly to a fully spin-polarized ground state at $\lambda\approx 5$ caused by the intra-dot Coulomb interaction ($z$-component of the total spin $S_z^{tot}$ changes from $1/2$ to $3/2$). This conclusion was based on results of QMC calculations (multilevel blocking algorithm). For the $1/2$ state calculations gave at $\lambda=4$ {\em slightly lower} energy, than for the $3/2$ state, and higher energies at $\lambda=6$, 8 and 10. In a subsequent (Erratum) publication \cite{Egger99b} other (corrected) values for the energy of these states were reported. At $\lambda=4$ they gave {\em the same} energies of the 1/2 and 3/2 states. The conclusion from these two publications in the part concerning the three-electron quantum dot is thus not clear. H\"ausler \cite{Hausler00}, calculating the energy spectrum of a three-electron dot using the so-called pocket state method, gave a more definite indication on the existence of such a transition. He presented his results, however, in some specific for the method and not-directly related to $\lambda$ units, which does not allow one to quantitatively characterize the transition (e.g. to get the value of the interaction parameter $\lambda$ at the transition point). H\"ausler also pointed out that some indications on such a transition can be also seen in earlier studies of a three-electron dot: in Ref. \cite{Hawrylak93} ($\lambda\approx 2$) the ground state of the dot at $B=0$ was found to be partly spin-polarized, while in Ref. \cite{Ruan95} with a substantially larger value of $\lambda$ the fully spin-polarized state turned out to have the lower energy. As seen from this brief outline, available data indicate that this transition seems to exist but it is not yet quantitatively understood, and its physical origin is not completely clear. It is also not clear whether and how this transition in the spin state of the system is related to the Fermi-liquid -- Wigner-molecule transition in the dot.

In this paper I present results of a complete theoretical study of a three-electron parabolic quantum dot. Using exact diagonalization technique, I calculate full energy spectrum of the dot, as a function of the interaction parameter in the range $0\le \lambda\le 10$. At $\lambda=\lambda_c=4.343$ I find a transition in the ground state of the dot, accompanied by the change of the total spin quantum number. I study the densities and the pair-correlation functions in the ground and the first excited states of the system at a number of $\lambda$-points, including the vicinity of the transition. These results show that physical properties of the dot dramatically change at the transition point, corresponding to properties of Fermi liquid at $\lambda<\lambda_c$, and of a Wigner molecule at $\lambda>\lambda_c$. I also calculate some thermodynamic properties of the dot: the heat capacity and the volume-pressure diagram. The temperature dependence of the heat capacity clearly exhibits characteristic features, related to the transition, in the temperature range corresponding to a few K for typical parameters of GaAs dots. Other experimental consequences from predictions of this paper are also discussed. 

In Section \ref{MMM} I briefly describe the model and the method used in calculations. Results of the work are presented in Section \ref{results}. Concluding remarks can be found in Section \ref{concl}. 

\section{Model and method}
\label{MMM}

\subsection{The Hamiltonian}

I consider three 2D electrons moving in the plane $z=0$ in a parabolic confining potential $V(r)=m^\star\omega_0^2 r^2/2$, ${\bf r}=(x,y)$. The Hamiltonian of the system
\begin{equation}
\hat H =\sum_{i=1}^N \left(\frac {\hat {\bf p}_i^2}{2m^\star} +
\frac{m^\star\omega_0^2 {\bf r}_i^2}2\right)
+ 
\frac 12\sum_{i\neq j=1}^N \frac{e^2}{
|{\bf r}_i-{\bf r}_j|} 
\label{qdhamiltonian}
\end{equation}
($N=3$) commutes with operators of the total angular momentum $\hat L_z^{tot}$, (squared) total spin $\hat {\bf S}_{tot}^2$ 
and projection of the total spin $\hat S_\zeta^{tot}$ on some ($\zeta$-) axis (not necessarily coinciding with the $z$-axis). This gives three conserving quantum numbers $L_{tot}\equiv L_z^{tot}$, $S_{tot}$ and $S_\zeta^{tot}$. No magnetic field is assumed to be applied to the system. 

\subsection{Basis set of single-particle states}

A complete set of single-particle solutions of the problem 
\begin{equation}
\phi_{nls}({\bf r},\sigma)=\varphi_{nl}({\bf r})\chi_s(\sigma)
\label{set}
\end{equation} 
is the product of the Fock-Darwin orbitals \cite{Fock28,Darwin31}
\begin{equation}
\varphi_{nl}({\bf r})= \frac{1}{l_0} \sqrt{\frac{n!}{\pi(n+|l|)!}} \left(\frac r{l_0}\right)^{|l|}e^{il\theta-r^2/2l_0^2}L_n^{|l|}(r^2/l_0^2)
\label{FD}
\end{equation}
and the spin functions $\chi_s(\sigma)$. Here $l_0=\sqrt{\hbar/m^\star\omega_0}$ is the oscillator length, and $(n,l,s)$ are the radial, asimutal (angular momentum) and spin quantum numbers of the single-particle problem ($n\ge 0$ and $l$ are integer, $s=\pm 1/2$). All the single-particle states (\ref{set}) can be ordered and enumerated, e.g. $\phi_1\equiv(nls)_1=(0,0,\uparrow)$, $\phi_2=(0,0,\downarrow)$, $\phi_3=(0,1,\uparrow)$, etc. The energy of the states (\ref{set}) does not depend on spins,
\begin{equation}
E_{nls}=\hbar\omega_0(2n+|l|+1).
\end{equation}

\subsection{Basis set of many-particle states}

A complete set of many-particle states $\Psi_u$, $u=1,2,\dots$ is formed by placing particles in different single-particle states, e.g. $\Psi_1=|\phi_1\phi_2\phi_3\rangle$, $\Psi_2=|\phi_1\phi_2\phi_4\rangle$, $\dots$, where $ |\phi_\alpha\phi_\beta\phi_\gamma\rangle$ are Slater determinants
\begin{equation}
|\phi_\alpha\phi_\beta\phi_\gamma\rangle=\frac 1{\sqrt{3!}}
\det\left|
\begin{array}{ccc}
\phi_\alpha(\xi_1) & 
\phi_\beta(\xi_1) &
\phi_\gamma(\xi_1) \\
\phi_\alpha(\xi_2) &
\phi_\beta(\xi_2) &
\phi_\gamma(\xi_2) \\
\phi_\alpha(\xi_3) &
\phi_\beta(\xi_3) &
\phi_\gamma(\xi_3) \\
\end{array}
\right|
\end{equation}
and $\xi_i=({\bf r}_i,\sigma_i)$. All the many-particle states $\Psi_u$ can be also arranged, e.g. in order of increasing of their total single-particle energy 
\begin{equation}
E_u^{sp}=E_{(nls)_\alpha}+E_{(nls)_\beta}+E_{(nls)_\gamma}, 
\label{totspenergy}
\end{equation}
and enumerated.

\subsection{Solution of the Schr\"odinger equation}

Expanding the many-body wave function in a complete set of many-particle states,
\begin{equation}
\Psi(\xi_1,\xi_2,\xi_3)=\sum_{u}
C_u\Psi_u(\xi_1,\xi_2,\xi_3),
\label{mbwf}
\end{equation}
I get the Schr\"odinger equation in the matrix form,
\begin{equation}
\sum_{u^\prime}
(H_{uu^\prime}-E\delta_{uu^\prime})C_{u^\prime}=0.
\label{evprob}
\end{equation}
The conservation of the total angular momentum $L_{tot}$ and the projection of the total spin $S_\zeta^{tot}$ allows one to chose the many-body states for the expansion (\ref{mbwf}) under additional constraints 
\begin{equation}
\sum_{i=1}^Nl_i=L_{tot},\ \ \sum_{i=1}^Ns_i=S_\zeta^{tot}.
\end{equation}
This reduces the size of the matrix in (\ref{evprob}) and facilitates calculations.

Numerically diagonalizing the eigenvalue problem (\ref{evprob}) I get a set of energy levels 
\begin{equation}
E_{L_{tot},S_\zeta^{tot},m}=\hbar\omega_0 {\cal F}_{L_{tot},S_\zeta^{tot},m}(\lambda),
\end{equation}
and corresponding eigenfunctions 
\begin{equation} 
\Psi_{L_{tot},S_\zeta^{tot},m}=\sum_uC_u^{L_{tot},S_\zeta^{tot},m}(\lambda)\Psi_u, 
\end{equation}
as a function of the interaction parameter $\lambda=l_0/a_B$. The number $m=1,2,\dots$ enumerates the energy levels of the system in the subspace of levels with given $L_{tot}$ and $S_\zeta^{tot}$. After the diagonalization problem is solved, the eigenvalues of the total spin are calculated for each level $m$ from
\begin{equation}
S_{tot}(S_{tot}+1)=
\langle \Psi_{L_{tot},S_\zeta^{tot},m}|\hat {\bf S}^2_{tot}|
\Psi_{L_{tot},S_\zeta^{tot},m}\rangle.
\end{equation}
All the matrix elements of the Hamiltonian $H_{uu^\prime}$ and of the operator $(\hat {\bf S}^2_{tot})_{uu^\prime}$ are calculated analytically. 

All the energy levels with non-zero $L_{tot}$ and $S_\zeta^{tot}$ are degenerate,
\begin{equation}
E_{L_{tot},S_\zeta^{tot}}=E_{-L_{tot},S_\zeta^{tot}}=E_{L_{tot},-S_\zeta^{tot}}=E_{-L_{tot},-S_\zeta^{tot}}.
\label{degeneracy}
\end{equation}
Presenting below results for the energy of the states $(L_{tot},S_\zeta^{tot})$, I omit the corresponding signs [for instance, $(1,1/2)$ stands for $(\pm 1,\pm 1/2)$ with all possible combinations of signs]. Degeneracy of levels are calculated accounting for (\ref{degeneracy}).

\subsection{Properties of the states and the heat capacity}

After the Schr\"odinger problem is solved and all the energy levels and the eigenfunctions are found, I calculate the density of spin-up and spin-down polarized electrons in the states $(L_{tot},S_\zeta^{tot},m)$, and the corresponding pair-correlation functions. These quantities are calculated as averages of the operators 
\begin{equation}
\hat n_\sigma({\bf r})=\sum_{i=1}^N\delta({\bf r}-{\bf r}_i)\delta_{\sigma\sigma_i}
\label{dens}
\end{equation}
and
\begin{equation}
\hat P_{\sigma\sigma^\prime}({\bf r},{\bf r}^\prime)=
\sum_{i=1}^N\sum_{j=1,\neq i}^N
\delta({\bf r}-{\bf r}_i)\delta({\bf r}^\prime-{\bf r}_j)
\delta_{\sigma\sigma_i}\delta_{\sigma^\prime\sigma_j}
\label{pcf}
\end{equation}
with the eigenfunctions $\Psi_{L_{tot},S_\zeta^{tot},m}$. All the matrix elements of the operators (\ref{dens}) and (\ref{pcf}) are calculated analytically.

As the method offers an opportunity to find all the energy levels of the system, one can also calculate thermodynamic properties of the dots. I calculate the heat capacity as $C_\lambda=(\partial\bar E/\partial T)_\lambda$, where
\begin{equation}
\bar E\equiv\bar E(\lambda,T)=\frac {\sum_n E_n(\lambda) e^{-E_n(\lambda)/T}}{\sum_n e^{-E_n(\lambda)/T}},
\end{equation} 
$T$ is the temperature, and the sum is taken over all (low-lying) energy levels accounting for their degeneracies. 

\subsection{Convergency of the method}

The number of all many-particle states in the problem is infinite, and the size of the matrix $H_{uu^\prime}$ in Eq. (\ref{evprob}) is infinite too. To perform practical calculations I restrict the number of many-particle states in the expansion (\ref{mbwf}) so that the total single-particle energy (\ref{totspenergy}) of the involved many-body states is smaller than some threshold value $E_{th}$, $E_u^{sp}\le E_{th}$. The larger the threshold energy $E_{th}$, the broader the range of $\lambda$ in which results are convergent and reliable. Typically, less than 1000 many-particle states were sufficient for all the calculations presented below.

\begin{figure}
\includegraphics[width=8.2cm]{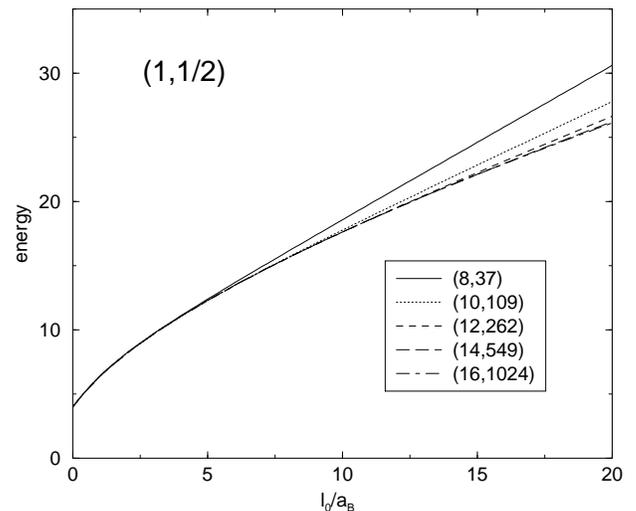}
\caption{Convergency of the energy of the lowest state with $(L_{tot},S_{tot})=(1,1/2)$. The energy unit is $\hbar\omega_0$. The curves are labeled by two numbers $(E_{th},N_{mbs})$, where $E_{th}$ is the threshold energy (in units $\hbar\omega_0$), and $N_{mbs}$ is the number of many-body states involved in the expansion (\ref{mbwf}).}
\label{convergency}
\end{figure}

Convergency of the method is illustrated on Figure \ref{convergency}, where the energy $E_{(1,1/2)}$ of the lowest state ($m=1$) with $(L_{tot},S_{tot})=(1,1/2)$ is shown as a function of $\lambda=l_0/a_B$ for increasing threshold energy $E_{th}$. The curves are labeled by $E_{th}$ and the number of many-body quantum states $N_{mbs}$ involved in the expansion (\ref{mbwf}). One sees that including about 1000 many-body states leads to very accurate results for the energy at $\lambda\le 20$. Notice that below I present results for the energy in the interval $\lambda\le 10$, where the method is practically exact: at $\lambda=10$ I found that $E_{(1,1/2)}/\hbar\omega_0=17.627891$ at $N_{mbs}=1024$, and $17.627974$ at $N_{mbs}=549$. The difference comprises $4.7\cdot 10^{-4}$ \%. 


\section{Results}
\label{results}

All the lengths in this Section are measured in units $l_0$, all the energies -- in units $\hbar\omega_0$, the densities and the pair-correlation functions -- in units $(\pi l_0^2)^{-1}$ and $(\pi l_0^2)^{-2}$ respectively. 

\subsection{Energy spectra}

\subsubsection{Exact results}

The interaction parameter in the problem
\begin{equation}
\lambda=\frac{l_0}{a_B}=\sqrt{\frac{e^2/a_B}{\hbar\omega_0}}\propto \frac{e^2}{\hbar^{3/2}}
\end{equation} 
characterizes the relative strength of classical Coulomb ($\sim e^2$) and quantization effects ($\sim\hbar$). The limit of small $\lambda$ corresponds to a weakly interacting system ($e^2\to 0$) and can be treated exactly. The ground state in this limit is realized in the configuration $[(0,0,\uparrow)(0,0,\downarrow)(0,1,\uparrow)]$, i.e.  $(L_{tot},S_{tot})=(1,1/2)$, with the energy 
\begin{equation}
\lim_{\lambda\to 0}E_{GS}/\hbar\omega_0 = 4.
\label{zerolambda}
\end{equation}

The opposite case $\lambda\to\infty$ corresponds to the purely classical limit ($\hbar\to 0$). In the classical ground state, electrons occupy the corners of an equilateral triangle \cite{Bolton93,Bedanov94}, with the distance 
\begin{equation}
R_{cl}=l_{cl}/3^{1/6}
\label{classradius}
\end{equation}
from the origin. The ground state energy at $\lambda=\infty$ is 
\begin{equation}
E^{cl}_{GS}=3^{5/3}\epsilon_{cl}/2.
\label{classGS}
\end{equation}
Here $l_{cl}=(e^2/m^\star\omega_0^2)^{1/3}
$ and $\epsilon_{cl}=e^2/l_{cl}$ are the classical length and energy units. 
%
So, in the classical limit 
\begin{equation}
\lim_{\lambda\to \infty}E_{GS}/\hbar\omega_0 = 3^{5/3} \lambda^{2/3}/2.
\label{infinitelmbda}
\end{equation}
Equations (\ref{zerolambda}) and (\ref{infinitelmbda}) give asymptotes of the ground state energy at very small and very large $\lambda$.

\begin{figure}
\includegraphics[width=8.2cm]{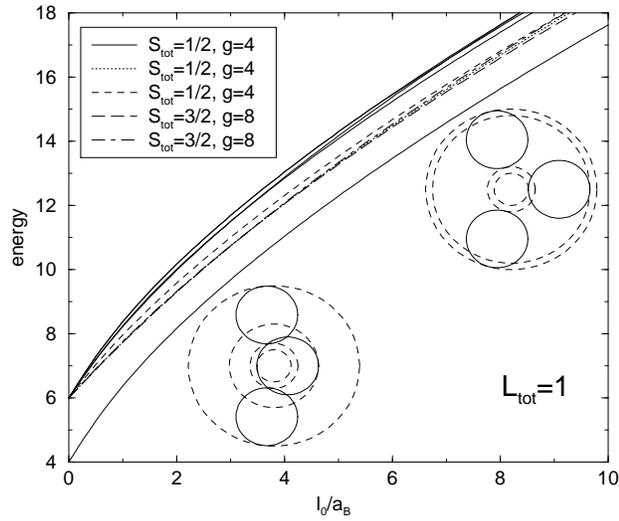}
\caption{Low-lying energy levels with the total angular momentum $L_{tot}=1$ in a 3-electron quantum dot. The energy unit is $\hbar\omega_0$. The five lowest levels are labeled by their total spin $S_{tot}$ and the degeneracy $g$. The insets schematically show the structure of the lowest-level wave function, at small (left inset) and large $\lambda$ (right inset), for details see Section \ref{sec:density}.}
\label{L1fullspectrum}
\end{figure}

At arbitrary $\lambda$ one needs numerical calculations. Figure \ref{L1fullspectrum} shows the low-lying energy levels of a three-electron parabolic quantum dot with the total angular momentum $L_{tot}=1$. Shown are only the states which start from $E/\hbar\omega_0\le 6$ at $\lambda=0$. The lowest-energy state (the ground state in the subspace of levels with $L_{tot}=1$) corresponds (in the limit $\lambda\to 0$) to the configuration $[(0,0,\uparrow)(0,0,\downarrow)(0,1,\uparrow)]$, and has the total spin $S_{tot}=1/2$. This state is four-fold degenerate ($L_{tot}=\pm 1/2$, $S_\zeta^{tot}=\pm 1/2$). 

\begin{figure}
\includegraphics[width=8.2cm]{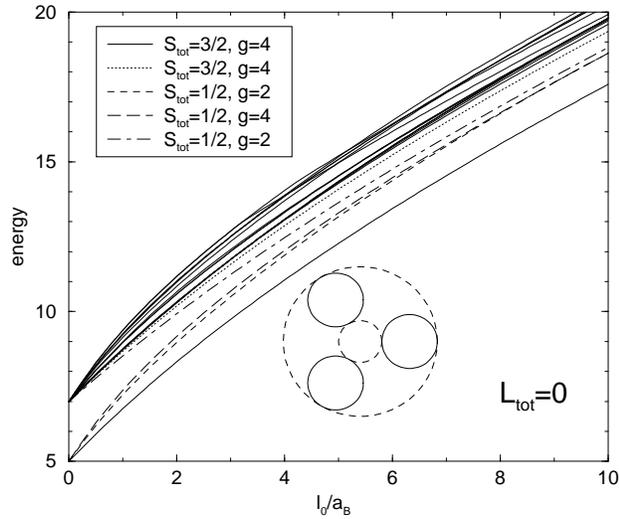}
\caption{Low-lying energy levels with the total angular momentum $L_{tot}=0$ in a 3-electron quantum dot. The energy unit is $\hbar\omega_0$. The five lowest levels are labeled by their total spin $S_{tot}$ and the degeneracy $g$. The inset schematically shows the structure of the lowest-level wave function, for details see Section \ref{sec:density}.}
\label{L0fullspectrum}
\end{figure}

Figure \ref{L0fullspectrum} shows the low-lying energy levels with the total angular momentum $L_{tot}=0$. Shown are only the states which start from $E/\hbar\omega_0\le 7$ at $\lambda=0$. The lowest-energy state has the total spin $S_{tot}=3/2$ and four-fold degeneracy ($S_\zeta^{tot}=\pm 3/2,\pm 1/2$). In the limit $\lambda\to 0$ the state with the full spin polarization $(S_{tot}=3/2,S_\zeta^{tot}=+3/2)$ corresponds to the configuration $[(0,-1,\uparrow)(0,0,\uparrow)(0,1,\uparrow)]$. The state with a partial spin polarization $(S_{tot}=3/2,S_\zeta^{tot}=+1/2)$ corresponds in the limit $\lambda\to 0$ to the configuration $(\Psi_1+\Psi_2+\Psi_3)/\sqrt{3}$, where
\begin{eqnarray}
\Psi_1&=&[(0,-1,\downarrow)(0,0,\uparrow)(0,1,\uparrow)], \nonumber \\
\Psi_2&=&[(0,-1,\uparrow)(0,0,\downarrow)(0,1,\uparrow)], \\
\Psi_3&=&[(0,-1,\uparrow)(0,0,\uparrow)(0,1,\downarrow)]. \nonumber 
\end{eqnarray}

\begin{figure}
\includegraphics[width=8.2cm]{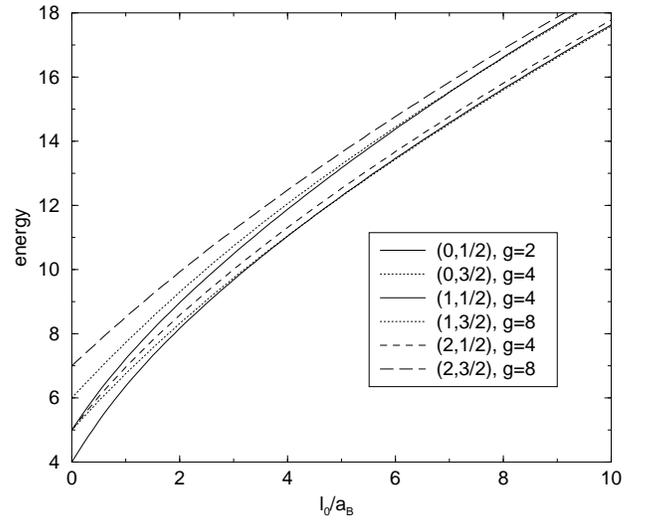}
\caption{Low-lying energy levels in a 3-electron quantum dot, for the total angular momentum $L_{tot}$ from 0 to 2 and for all total spin states. Only the one lowest-energy level is shown for each pair of quantum numbers $(L_{tot},S_{tot})$. The energy unit is $\hbar\omega_0$. The levels are labeled by the pair of quantum numbers $(L_{tot},S_{tot})$ and the degeneracy $g$.}
\label{allLlowlevels}
\end{figure}

\begin{figure}
\includegraphics[width=8.2cm]{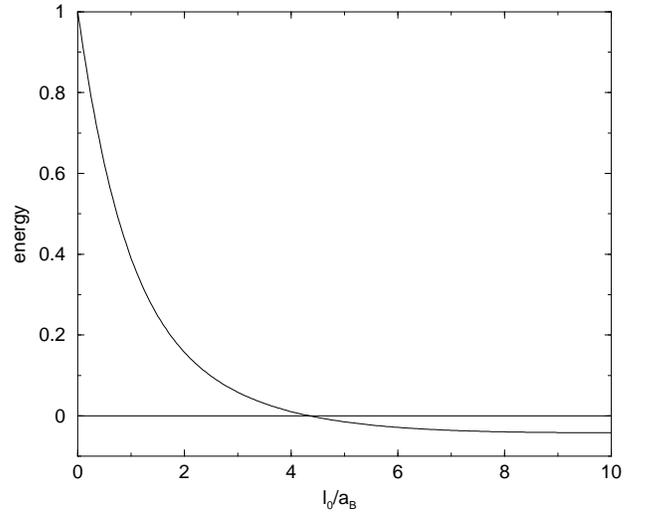}
\caption{Energy difference between the lowest states $E_{(0,3/2)}-E_{(1,1/2)}$  in a 3-electron quantum dot as a function of the interaction parameter $\lambda=l_0/a_B$. The energy unit is $\hbar\omega_0$. The transition occurs at $\lambda=\lambda_c\approx 4.343$.}
\label{gs-diff}
\end{figure}

I performed similar calculations for $L_{tot}$ from 0 to 9 and for all total-spin states $S_{tot}=3/2$ and 1/2. Results for $L_{tot}$ from 0 to 2 are shown in Figure \ref{allLlowlevels}. {\em Only the lowest-energy levels} ($m=1$) {\em for each pair of numbers $(L_{tot},S_{tot})$ are shown in the Figure} [this means that, say, a (1,1/2)-level with $m>1$ (not shown in the Figure) may have the lower energy than the exhibited level $(2,3/2)$ with $m=1$]. At some critical value of $\lambda$ ($\lambda=\lambda_c= 4.343$) one observes a {\em crossing} of the two lowest-energy levels $(1,1/2)$ and $(0,3/2)$ (more clearly seen in Figure \ref{gs-diff} where the energy difference $E_{(0,3/2)}-E_{(1,1/2)}$ is plotted versus $\lambda$). At the critical point the total spin of the system in the ground state changes from $S_{tot}=1/2$ at $\lambda<\lambda_c$ to $S_{tot}=3/2$ at $\lambda>\lambda_c$. Near the critical point the gap between the ground and the first excited states behaves as

\begin{equation}
|E_{(0,3/2)}-E_{(1,1/2)}|/\hbar\omega_0=0.02766|\lambda-\lambda_c|,
\end{equation}
with a jump of the derivative of the ground state energy with respect to $\lambda$. At large $\lambda$ (more exactly, at $\lambda=10$) the energy difference is $E_{(0,3/2)}-E_{(1,1/2)}\approx -0.0416 \hbar\omega_0$.

In Table \ref{tab1} exact results for the energies of the states (1,1/2) and (0,3/2) are compared with QMC results from Ref. \cite{Egger99b}. One sees that the accuracy of the QMC results \cite{Egger99b} is in general very high, but the errors are not always small (see e.g. $\lambda=10$) compared to the {\em difference} between the energies of the ground and excited states. 

\subsubsection{Approximations}

The energy of the states $(1,1/2)$ and $(0,3/2)$ can be approximated, in the interval $0\le\lambda\le 10$, by the formula
\begin{equation}
E=E^{cl}_{GS}+\epsilon_{cl}\left(AX + \sqrt{B^2 + C^2 X^2} -B\right),
\label{approx11}
\end{equation}
where $E^{cl}_{GS}$ is the classical ground state energy (\ref{classGS}), $X=\hbar\omega_0/\epsilon_{cl}$, and the numbers $A,B,C$ for the two considered states are 
\begin{equation}
(A,B,C)_{(1,1/2)}=(3.11536,2.93076,0.917954), 
\label{numbers11}
\end{equation}
\begin{equation}
(A,B,C)_{(0,3/2)}=(2.84171,2.44529,2.13633). 
\label{numbers03}
\end{equation}
For the state $(1,1/2)$ the difference between the exact energy and the approximation (\ref{approx11})--(\ref{numbers11}) is about 0.83\% at $\lambda=0$, does not exceed 0.22\% at $0.05\le \lambda\le 10$, and tends to zero with growing $\lambda$. For the state $(0,3/2)$ the difference between the exact solution and the approximation is smaller than 0.44\% in the whole range of $\lambda$. It should be noted however, that the difference between the energies of the two states $E_{(0,3/2)}-E_{(1,1/2)}$ is reproduced by the approximate formulas (\ref{approx11})--(\ref{numbers03}) with a substantially worse accuracy.

\subsection{Heat capacity}
\label{sec:heatcap}

\begin{figure}
\includegraphics[width=8.2cm]{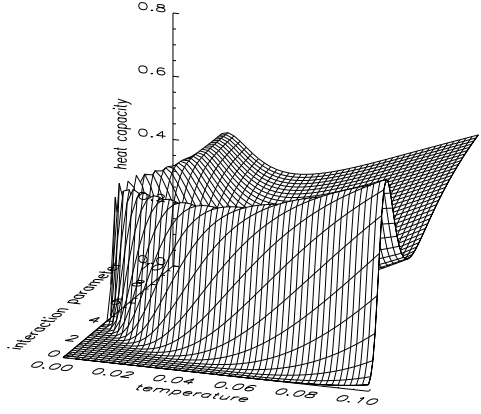}
\caption{Heat capacity of a 3-electron parabolic quantum dot, as a function of the temperature $kT/\hbar\omega_0$ and the interaction parameter $l_0/a_B$. }
\label{heatcapacity}
\end{figure}

Figure \ref{heatcapacity} exhibits the calculated low-temperature heat capacity $C_\lambda$ as a function of $T$ and $\lambda$. About 30 lowest-energy levels for each $L_{tot}$ from 0 to 9, with corresponding degeneracies, were taken into account in this calculation. A pronounced peak related to the gap between the ground and the first excited state can be clearly seen in this Figure. The value of the peak temperature $T_p(\lambda)$ depends on $\lambda$ as $|E_{(0,3/2)}-E_{(1,1/2)}|$ (compare to Figure \ref{gs-diff}), and disappears at the critical point $\lambda=\lambda_c$. The most dramatic variations of the heat capacity are the case in the range $kT\lesssim 0.1 \hbar\omega_0$, which corresponds, at a typical confinement of GaAs quantum dots ($\hbar\omega_0\sim 3$ meV), to a few-K temperature scale.


\subsection{Electron density and correlations}
\label{sec:density}

Due to rotational symmetry of the Hamiltonian (\ref{qdhamiltonian}) the density $n_\sigma^{(L_{tot},S_{tot})}(r,\theta)$ of spin-up and spin-down polarized electrons in the quantum-mechanical states $(L_{tot},S_{tot})$ does not depend on the angular coordinate $\theta$, and are shown below as a function of $r$ only. The pair-correlation functions $ P_{\sigma\sigma^\prime}({\bf r},{\bf r}^\prime)$ are plotted as a function of ${\bf r}/l_0$, for all orientations of spins, at the second coordinate ${\bf r}^\prime$ being fixed at the classical distance (\ref{classradius}) from the origin, ${\bf r}^\prime=(0,-R_{cl})$ (the {\em second} subscript corresponds to the spin of a fixed electron).

\subsubsection{The state $(L_{tot},S_{tot})=(0,3/2)$}

\begin{figure}
\includegraphics[width=8.2cm]{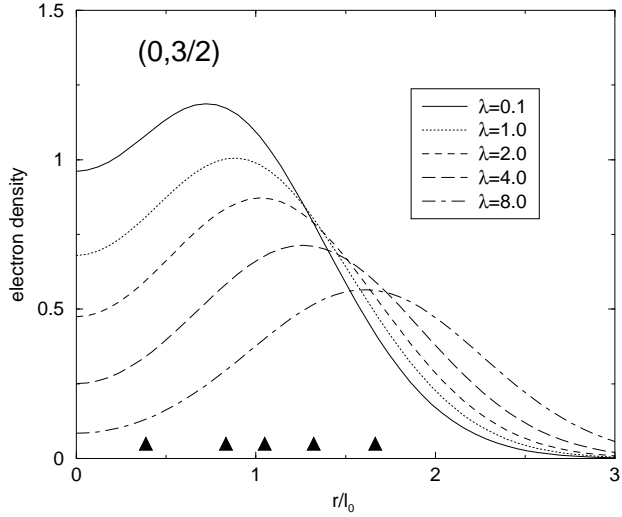}
\caption{Total electron density (spin-up plus spin-down) in the state $(L_{tot},S_{tot})=(0,3/2)$ at different $\lambda$. Triangles show the positions of the classical radius  (\ref{classradius}) at corresponding values of $\lambda$.}
\label{dens03}
\end{figure}

The total density of electrons $n^{(0,3/2)}(r)=n_\uparrow^{(0,3/2)} (r)+n_\downarrow^{(0,3/2)} (r)$ in the state $(L_{tot},S_{tot})=(0,3/2)$ at a few values of the interaction parameter $\lambda$ is shown in Figure \ref{dens03}. The function $n^{(0,3/2)}(r)$ also determines the densities of spin-up and spin-down polarized electrons: in the state with the total spin projection $S_\zeta^{tot}=+3/2$ the density of spin-up electrons coincides with $n^{(0,3/2)}(r)$, while the density of spin-down electrons is zero; in the $S_\zeta^{tot}=+1/2$ state one has $n_\uparrow^{(0,3/2,+1/2)}(r)=\frac 23n^{(0,3/2)}(r)$, and $n_\downarrow^{(0,3/2,+1/2)}(r)=\frac 13n^{(0,3/2)}(r)$. One sees that at all $\lambda$ maxima of the electron density lie at a finite distance from the origin. At $\lambda\gtrsim 1$ they are very close to the classical radii (\ref{classradius}), shown by triangles on the Figure; at $\lambda\ll 1$ (weak Coulomb repulsion) they are at even larger $r$, due to the exchange repulsion.

\begin{figure*}
\includegraphics[width=8.2cm]{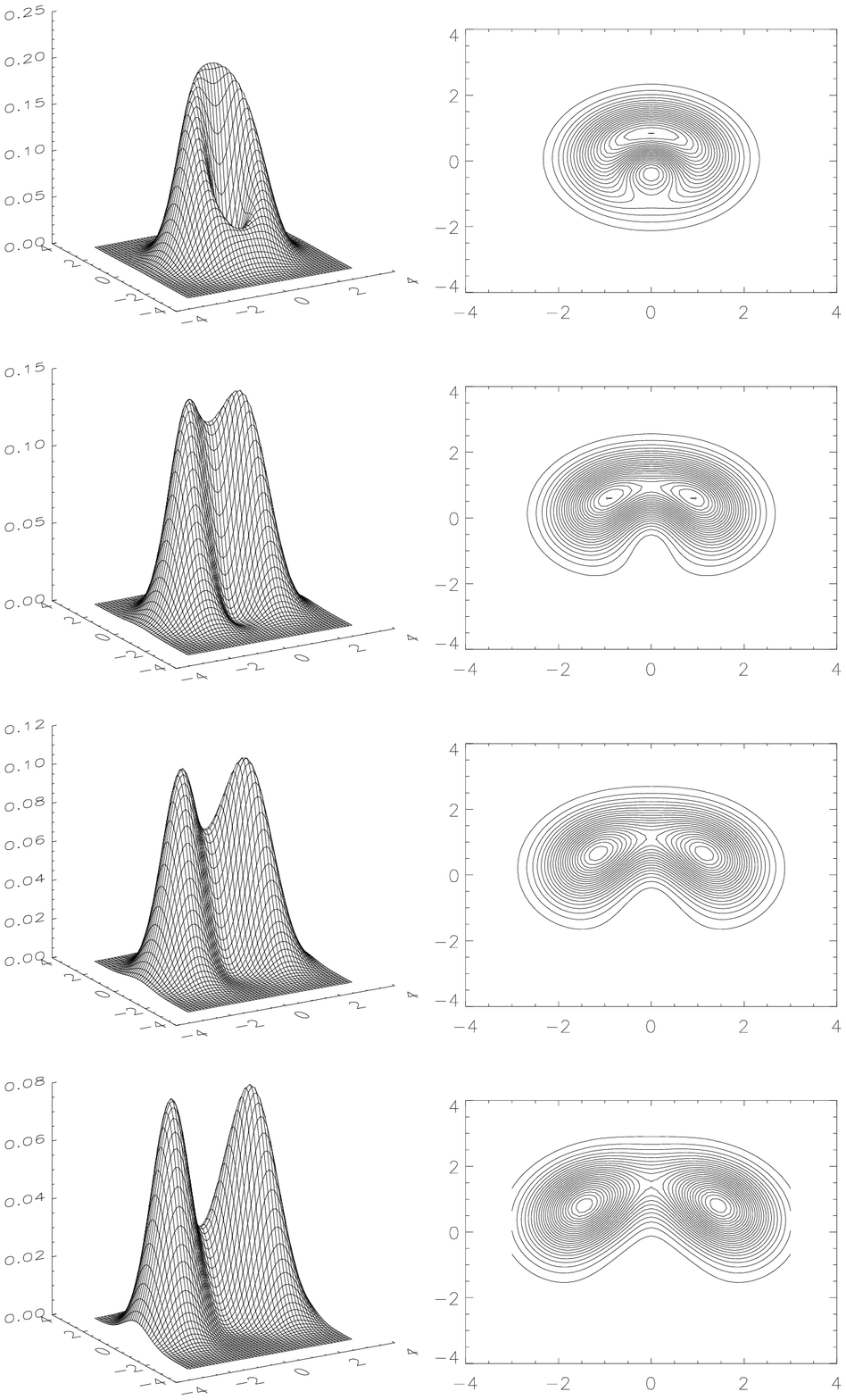}
\caption{Pair-correlation function $P_{\uparrow\uparrow}$ of the state with $L_{tot}=0$, $S_{tot}=3/2$, $S_\zeta^{tot}=+3/2$, at $\lambda=0.1$, 
2.0, 4.0, and 8.0, from up to down ($|{\bf r}^\prime|/l_0=R_{cl}/l_0=0.38$, 1.04, 1.32, and 1.66, respectively).}
\label{pcf033}
\end{figure*}

Figure \ref{pcf033} exhibits the pair-correlation function $P_{\uparrow\uparrow}({\bf r},{\bf r}^\prime)$ in the state $(L_{tot},S_{tot},S_\zeta^{tot})=(0,3/2,+3/2)$ (three other functions in this state obviously vanish, $P_{\uparrow\downarrow}=P_{\downarrow\uparrow}=P_{\downarrow\downarrow}=0$; in the state with $S_\zeta^{tot}=+1/2$ one has $P_{\uparrow\uparrow,\uparrow\downarrow,\downarrow\uparrow}^{(0,3/2,+1/2)}=\frac 13P_{\uparrow\uparrow}^{(0,3/2,+3/2)}$, and $P_{\downarrow\downarrow}^{(0,3/2,+1/2)}=0$). The interaction parameter $\lambda$ assumes the values 0.1, 2, 4, and 8, from up to down. At small $\lambda$ electron-electron interaction is weak, and the pair-correlation function has a form of a single peak centered opposite to the fixed electron. With growing $\lambda$ this peak is splitted onto two ones, and this splitting becomes very pronounced at strong Coulomb interaction ($\lambda\gtrsim 4$). 

The internal structure of the state $(0,3/2)$ thus corresponds to the angle-averaged classical configuration of an {\em equilateral} triangle (see the inset to Figure \ref{L0fullspectrum}). This structure is highly symmetric and ``rigid'': the ratio of the sides of the triangle remains unchanged when the curvature $K$ of the confiment potential varies ($K\propto\lambda^{-4}$). This state is a quantum-dot analog of the Wigner solid (a Wigner molecule).

\subsubsection{The state $(L_{tot},S_{tot})=(1,1/2)$}

\begin{figure}
\includegraphics[width=8.2cm]{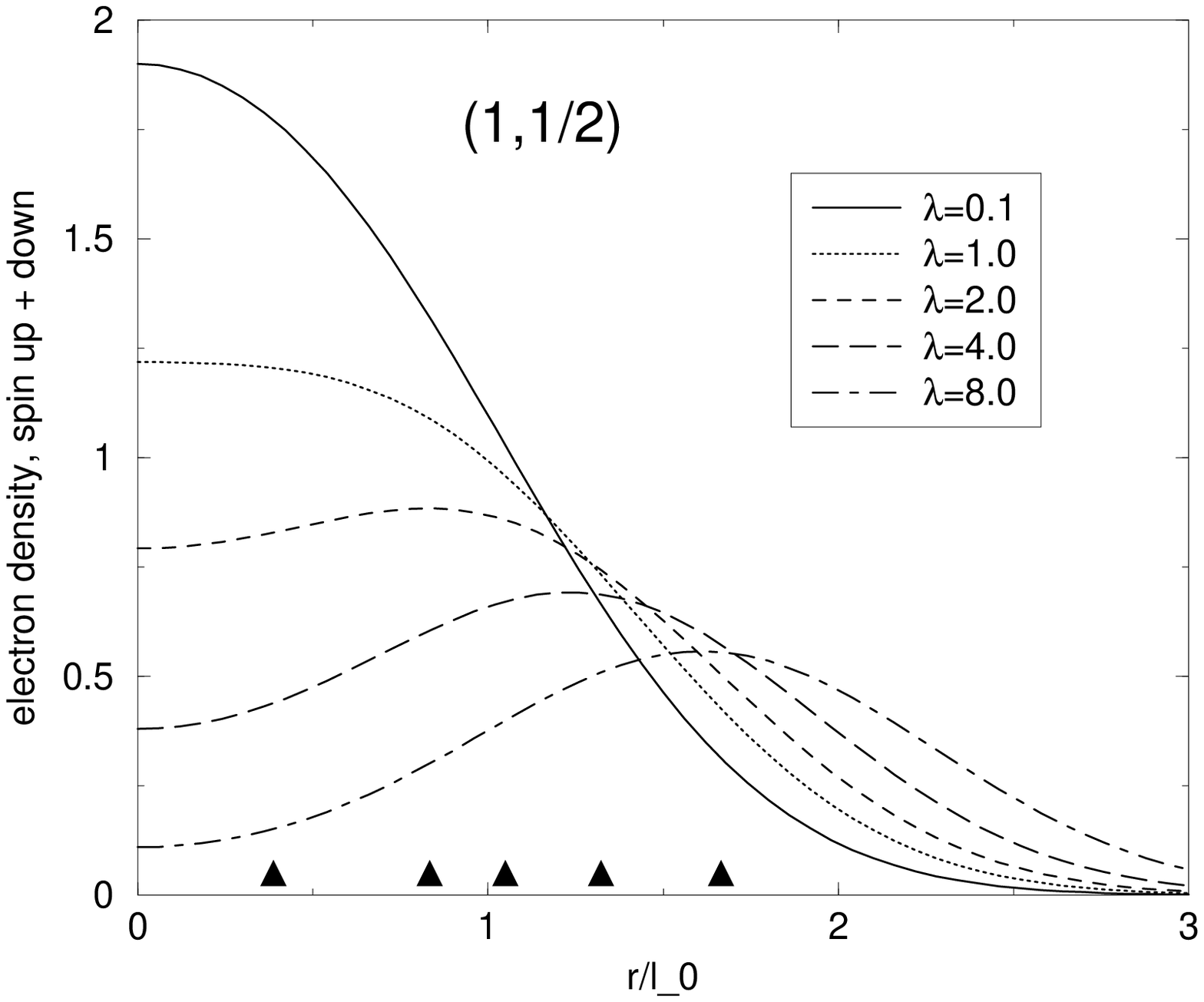}
\caption{Total electron density (spin-up plus spin-down) in the state (1,1/2) at different $\lambda$. Triangles show the positions of the classical radius  (\ref{classradius}) at corresponding values of $\lambda$. }
\label{dens11}
\end{figure}

In the state $(1,1/2)$ the spatial distribution of electrons is less trivial. Now we have two sorts of particles: two electrons are polarized up, and one electron is polarized down. Figure \ref{dens11} exhibits the total electron density (spin-up plus spin-down). One sees that at small $\lambda$ electrons behave as non- or weakly interacting particles (a Fermi-liquid-type state), forming the structure with maximum of the electron density at $r=0$. Such a picture is the case up to $\lambda\simeq 2$, when a weak minimum of the density at $r=0$ appears. At even larger $\lambda$ the influence of electron-electron interaction becomes more important: the density of electrons qualitatively looks like in the state (0,3/2), with a minimum at $r=0$ and maxima close to the classical radii (\ref{classradius}). 

\begin{figure}
\includegraphics[width=8.2cm]{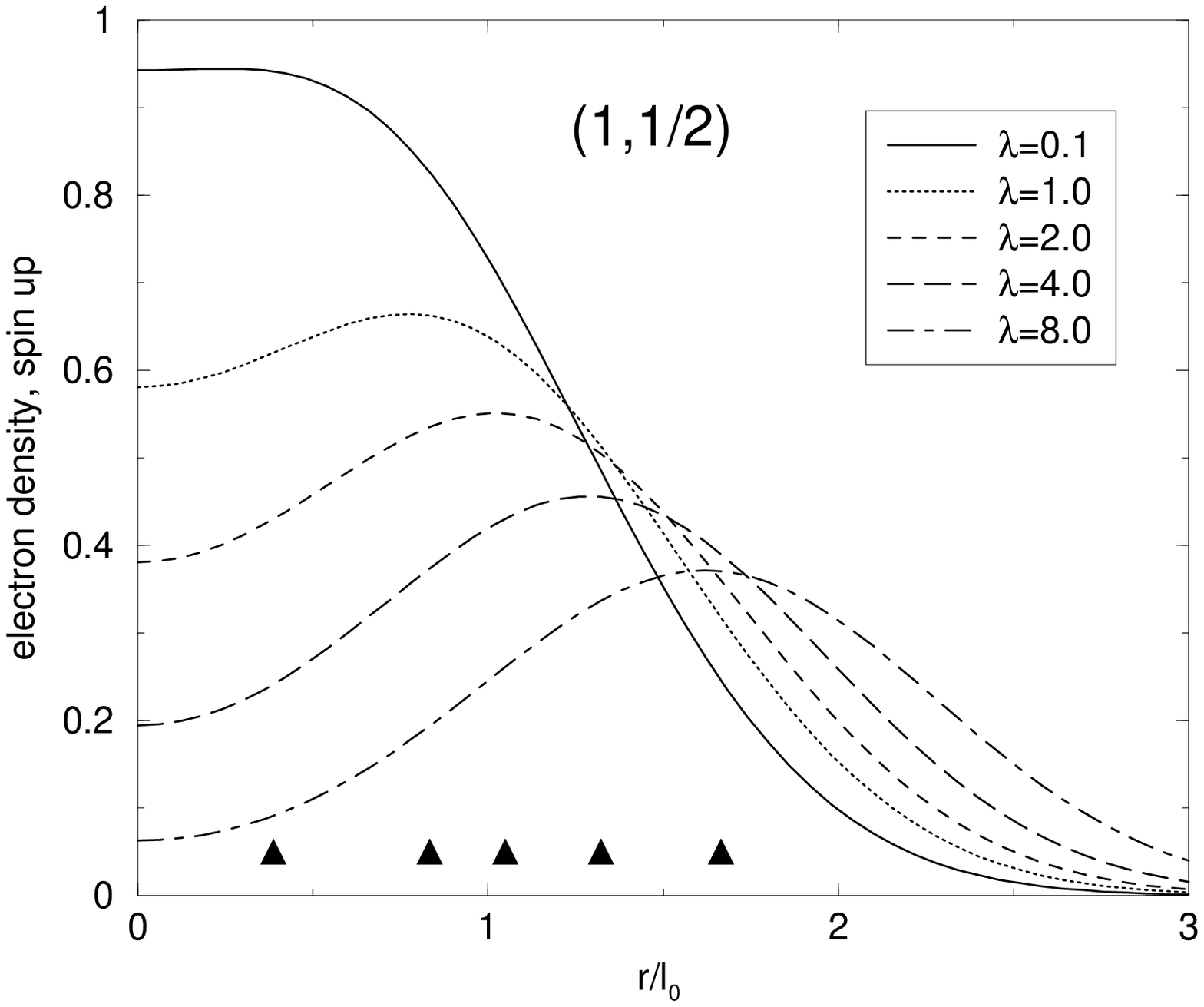}
\caption{Density of spin-up polarized electrons in the state (1,1/2) at different $\lambda$. Triangles show the positions of the classical radius  (\ref{classradius}) at corresponding values of $\lambda$.}
\label{dens11U}
\end{figure}

\begin{figure}
\includegraphics[width=8.2cm]{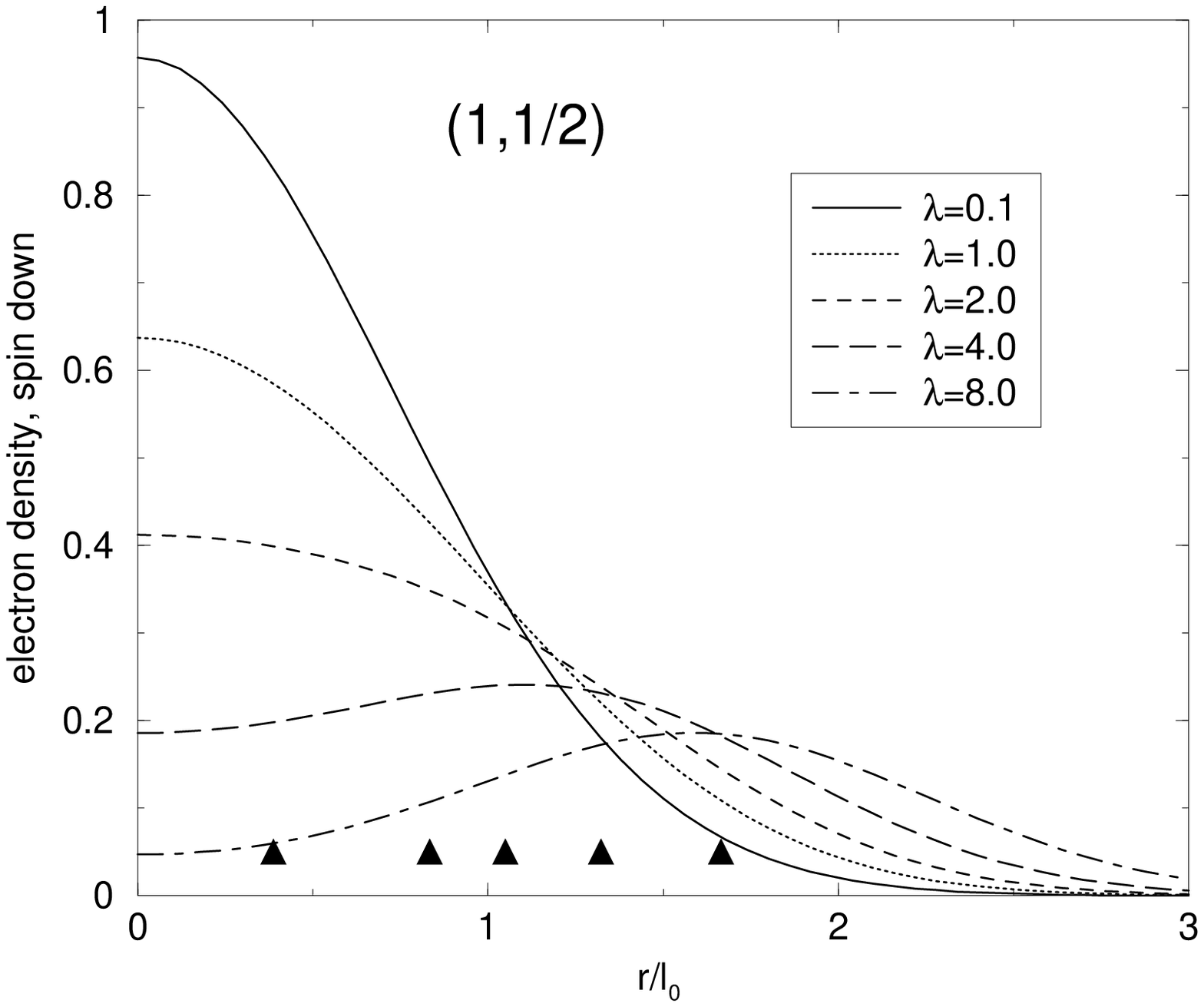}
\caption{Density of spin-down polarized electrons in the state (1,1/2) at different $\lambda$. Triangles show the positions of the classical radius  (\ref{classradius}) at corresponding values of $\lambda$.}
\label{dens11D}
\end{figure}

Additional and even more interesting information can be extracted from Figures \ref{dens11U} and \ref{dens11D}, which show separately the densities of spin-up and spin-down polarized electrons. One sees that at small $\lambda$ the one spin-down electron occupies the center of the dot, while the two spin-up electrons rotate around the center with a maximum of the density at a finite distance from the origin. Such a situation is a peculiar quantum-mechanical feature: it is not encountered in the classical picture. With the growing $\lambda$ (electron-electron interaction increases) the two spin-up electrons push the spin-down electron out from the center, but the structure ``one (spin-down) electron is essentially closer to the center, two (spin-up) electrons rotate around'' conserves up to $\lambda\simeq 2$: the density of the spin-up electrons has a clear maximum at a finite distance from the origin, while the density of spin-down electron still maximal at $r=0$. At even larger $\lambda$ ($\gtrsim 4$) (stronger electron-electron interaction) the density maximum of the spin-down electron is shifted to a finite $r$, but at any $\lambda$ it is closer to the origin,
than the density maximum of the spin-up electrons.

\begin{figure*}
\includegraphics[width=8.2cm]{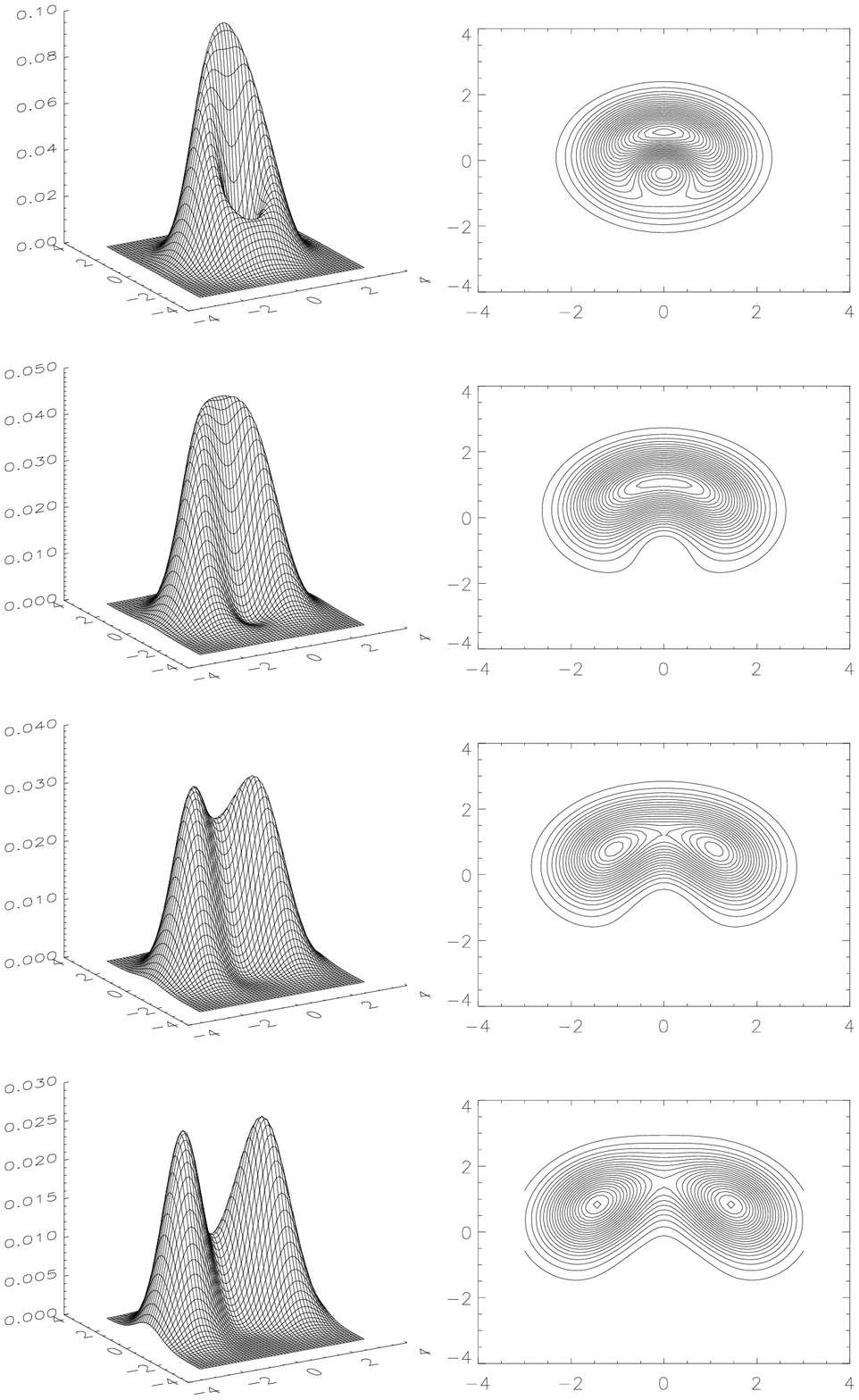}
\caption{Pair-correlation function $P_{\uparrow\uparrow}$ of the state with $L_{tot}=1$, $S_{tot}=1/2$, $S_\zeta^{tot}=+1/2$, at $\lambda=0.1$, 2, 4, and 8, from up to down ($|{\bf r}^\prime|/l_0=R_{cl}/l_0=0.38$, 1.04, 1.32, and 1.66, respectively).}
\label{pcf111a}
\end{figure*}

\begin{figure*}
\includegraphics[width=8.2cm]{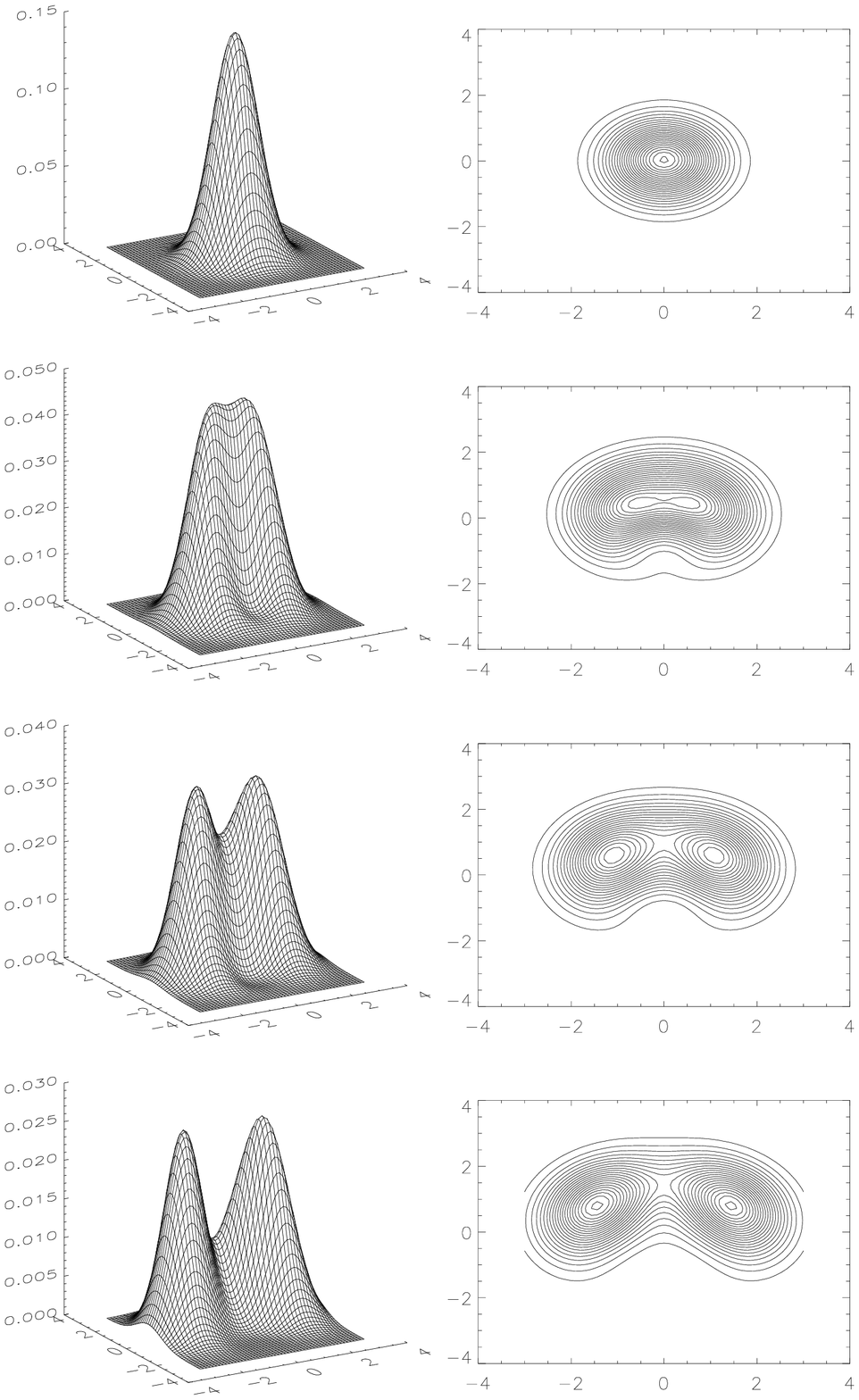}
\caption{Pair-correlation function $P_{\downarrow\uparrow}$ of the state with $L_{tot}=1$, $S_{tot}=1/2$, $S_\zeta^{tot}=+1/2$, at $\lambda=0.1$, 2, 4, and 8, from up to down ($|{\bf r}^\prime|/l_0=R_{cl}/l_0=0.38$, 1.04, 1.32, and 1.66, respectively).}
\label{pcf11b}
\end{figure*}

\begin{figure*}
\includegraphics[width=8.2cm]{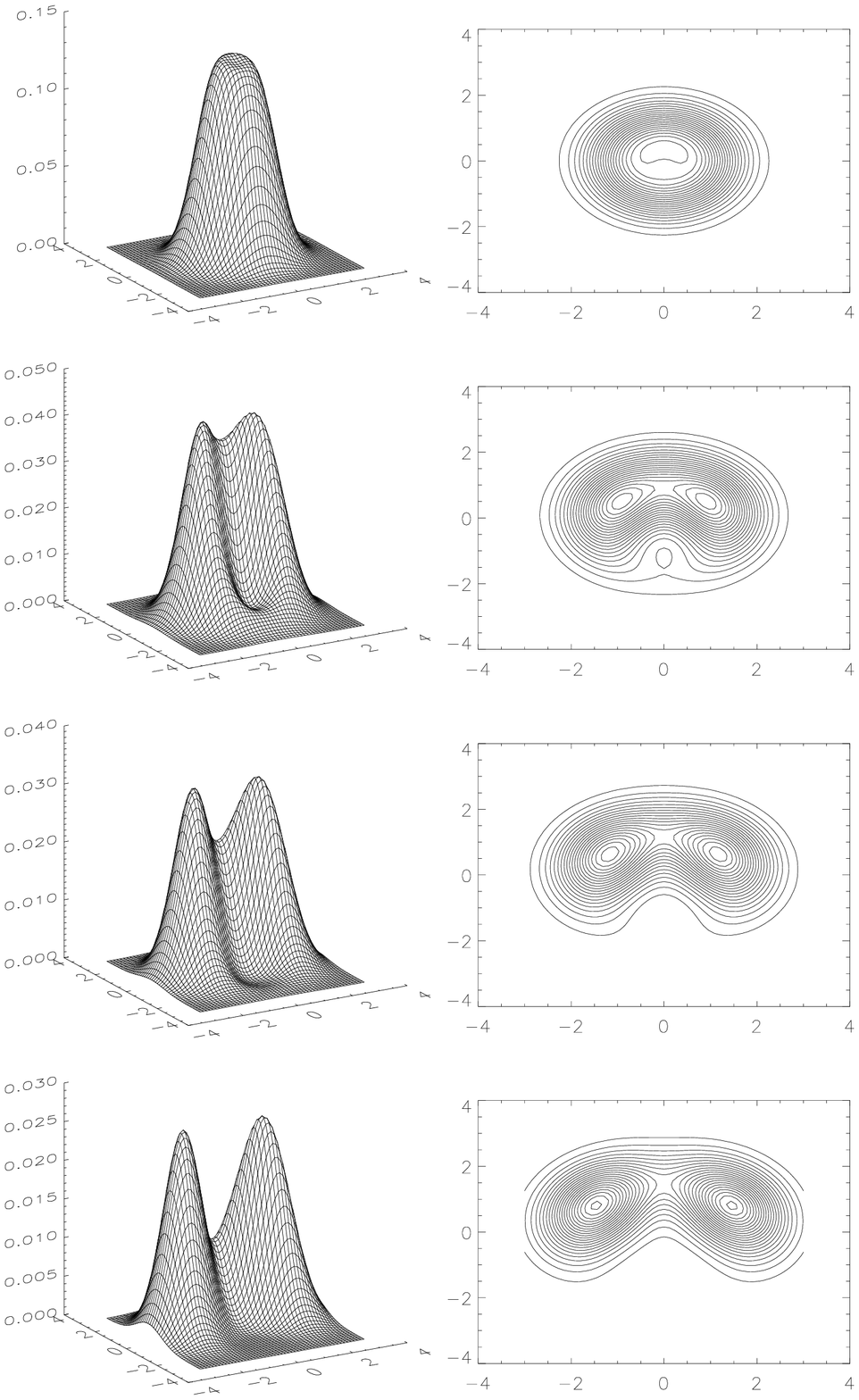}
\caption{Pair-correlation function $P_{\uparrow\downarrow}$ of the state with $L_{tot}=1$, $S_{tot}=1/2$, $S_\zeta^{tot}=+1/2$, at $\lambda=0.1$, 2, 4, and 8, from up to down ($|{\bf r}^\prime|/l_0=R_{cl}/l_0=0.38$, 1.04, 1.32, and 1.66, respectively).}
\label{pcf11c}
\end{figure*}

These features can be also seen in Figures \ref{pcf111a}, \ref{pcf11b}, and \ref{pcf11c}, which exhibit the pair-correlation functions $P_{\uparrow\uparrow}$, $P_{\downarrow\uparrow}$, and $P_{\uparrow\downarrow}$ in the state $(L_{tot},S_{tot},S_\zeta^{tot})=(1,1/2,+1/2)$ ($P_{\downarrow\downarrow}=0$ in this state). Compare for instance Figures \ref{pcf111a} ($P_{\uparrow\uparrow}$) and \ref{pcf11b} ($P_{\downarrow\uparrow}$). In both cases a spin-up electron is fixed at the classical distance from the origin. Let $\lambda=2$ (the second row of plots). One sees that the maximum of the pair-correlation function is about two times closer to the center of the dot for the spin-down electron (Figure \ref{pcf11b}) than for the spin-up electron (Figure \ref{pcf111a}). At $\lambda=4$ this difference is less pronounced but can also be seen. In Figure \ref{pcf11b} one also sees that in the limit of weak Coulomb interaction $\lambda\ll 1$ the spin-down electron is localized in the center, in agreement with the above analysis of the density plots.

The internal structure of the state $(1,1/2)$ thus corresponds to an angle-averaged configuration of an {\em isosceles} triangle, with two spin-up electrons at the base corners and one spin-down electron at the top of the triangle. This structure is less symmetric than that of the $(0,3/2)$ state and ``soft'': the ratio of the sides varies with $\lambda$, changing from 1/2 at $\lambda=0$ to 1 at $\lambda=\infty$ (see the insets to Figure \ref{L1fullspectrum}). This state is of a Fermi-liquid type (Fermi-gas at $\lambda\to 0$).

\subsubsection{Reconstruction of the ground state}

Now, we can understand the physical reason of the transition $(1,1/2)\leftrightarrow(0,3/2)$ at the varying parameter $\lambda$. Consider what happens with the ground state of the system, when the curvature $K\propto \omega_0^2\propto\lambda^{-4}$ of the confinement potential varies from small (the limit of strong Coulomb interaction $\lambda\to\infty$) to large values (the weak Coulomb interaction regime). At small $K$ the quantization effects are negligible, $\hbar\omega_0/\epsilon_{cl}=\lambda^{-2/3}\ll 1$, electrons are at a large (compared to $l_0$) distance from each other, and form a quasi-classical equilateral-triangle structure. In this highly-symmetric structure all electrons should be equivalent (have the same spin), therefore the total spin of the dot in this limit is $S_{tot}=3/2$. Increasing the curvature reduces the area of the triangle. Its form however first remains unchanged. Further increase of the curvature costs energy, and the system is reconstructed, at $\lambda=\lambda_c$, to another ground state with a more compact, ``soft'' isosceles-triangle structure. In this less symmetric structure one electron should differ from two others, therefore the transition to the new ground state is accompanied by the change of the total spin $S_{tot}\to 1/2$. Further increase of the curvature changes the ratio of sides of the isosceles triangle, but not its symmetry.

\begin{figure}
\includegraphics[width=8.2cm]{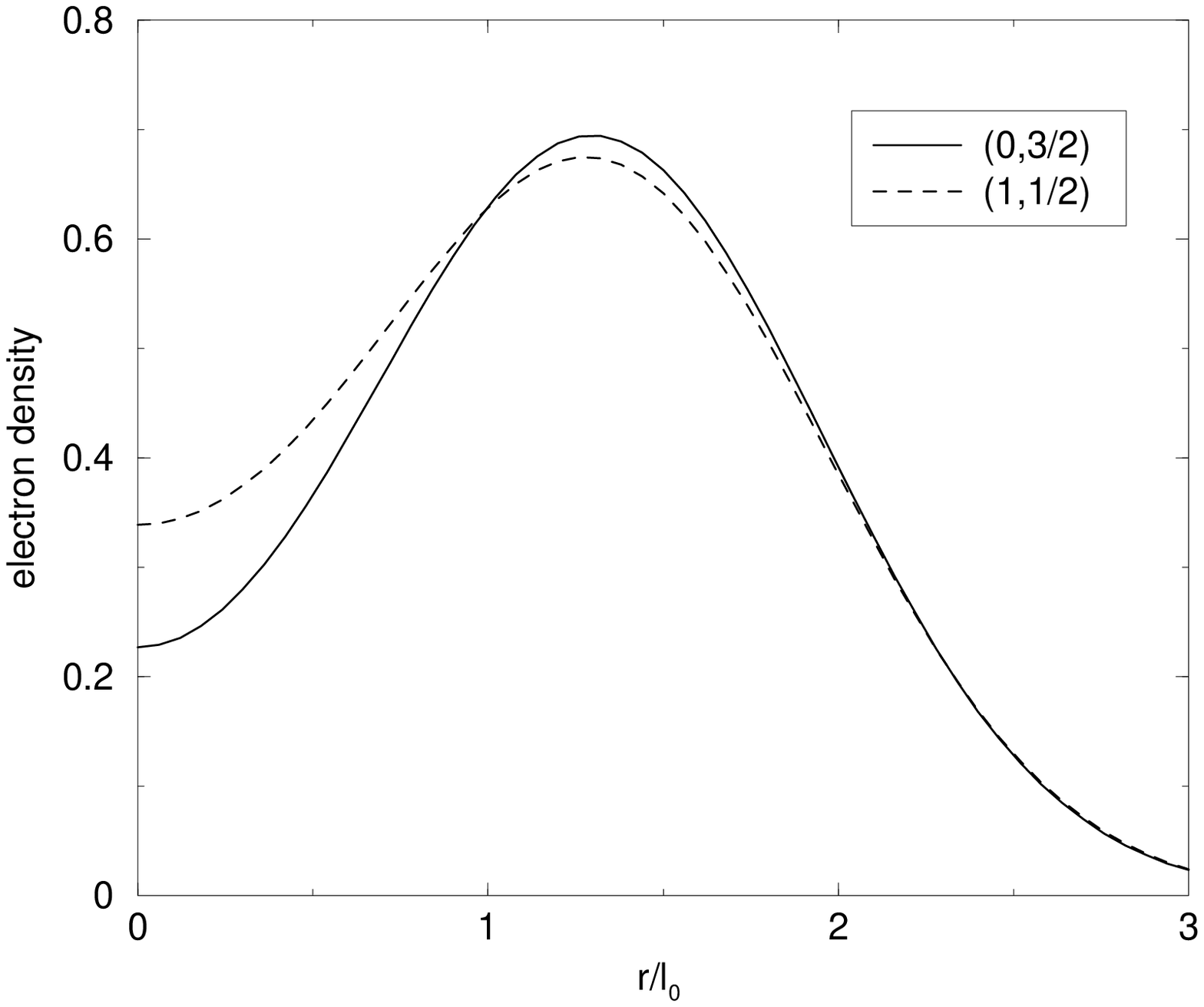}
\caption{Electron density in the states (1,1/2) (the ground state at $\lambda<\lambda_c$) and (0,3/2) (the ground state at $\lambda>\lambda_c$) at the transition point $\lambda=\lambda_c=4.343$.}
\label{denstrans}
\end{figure}

\begin{figure}
\includegraphics[width=8.2cm]{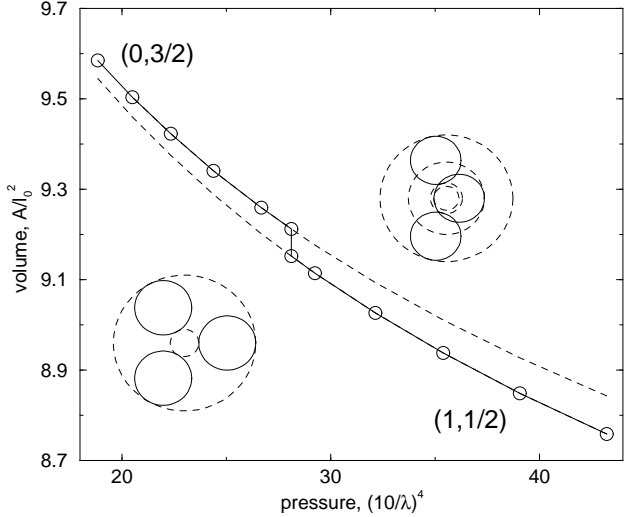}
\caption{A ``volume''-``pressure'' diagram, $A/l_0^2$ versus $(10/\lambda)^4$, for the ground state of a quantum dot lithium in the vicinity of the transition point $\lambda=\lambda_c=4.343$ (solid curve). Dashed curves show the ``volume''-``pressure'' diagrams for the states $(0,3/2)$ (upper curve) and $(1,1/2)$ (lower curve). Insets schematically show the distribution of electrons in the ground state on both sides of the transition point. In a real system the solid curve may have a hysteresis.}
\label{VPcurve}
\end{figure}

Exactly at the transition point $\lambda=\lambda_c$ the physical properties of the dot change very sharply. Figure \ref{denstrans} shows the total density of electrons $n_e(r)$ at $\lambda=\lambda_c$ in the states $(0,3/2)$ and $(1,1/2)$. When the system passes from the $(0,3/2)$ to the (1,1/2) state, electrons are pushed towards the center of the dot with a 50\% increase of the density at the point $r=0$. Figure \ref{VPcurve} shows the area of the dot 
 
\begin{equation}
A=\frac 1N \int d{\bf r}\pi r^2n_e(r)=\frac \pi N \langle\Psi_{GS}|\sum_i r_i^2|\Psi_{GS}\rangle
\end{equation}
[$\Psi_{GS}$ is the ground-state wave function] versus the curvature $K\propto\lambda^{-4}$ near the critical point $\lambda=\lambda_c$. As the curvature of the confinement potential can be treated as a ``pressure'' acting on electrons of the dot from the confining potential, Figure \ref{VPcurve} can be considered as a ``volume''--``pressure'' diagram. One sees that increasing the pressure leads to a discontinuous jump of the volume (with a $\delta$-like peculiarity in the compressibility) at the critical point $\lambda=\lambda_c$. In a real system the transition shown in Figure \ref{VPcurve} may happen with a hysteresis.

\section{Concluding remarks}
\label{concl}

The transition $(0,3/2)\leftrightarrow (1,1/2)$ in the quantum-dot lithium is very similar to the Fermi-liquid -- Wigner-solid transition in an infinite 2DES \cite{Tanatar89}. This becomes evident if to plot results of quantum-dot and 2DES calculations in the form of an expansion in powers of $\hbar$: energy in units $\epsilon_{cl}$ vs $\lambda^{-2/3}$ in dots, and energy in units $e^2\sqrt{n_s}$ vs $r_s^{-1/2}$ in 2DES; here $r_s=1/\sqrt{\pi n_s}a_B$. In such coordinates the ground-state-energy curves, both in dots and in 2DES, look very similarly: the Fermi-liquid and the Wigner-solid energies are smooth, almost linear functions [for dots see Eq. (\ref{approx11})], intersecting each other at some value of the interaction parameter. In the regime of strong interaction the fully spin-polarized Wigner solid is the ground state in both the 2DES and the dot; in the regime of weak interaction the unpolarized (partly-polarized in three-electron dots) Fermi-liquid state is the ground state of both systems. The transition $(0,3/2)\leftrightarrow (1,1/2)$ in the quantum-dot lithium is thus an analog (or a precursor) of the Fermi-liquid--Wigner-solid transition in an infinite 2DES \cite{Tanatar89}. 

The transition $(1,1/2)\leftrightarrow (0,3/2)$ should be observable in many experiments. It should manifest itself in any thermodynamic quantity. The difference in the ground-state total spin $S_{tot}$ should be also seen in the Zeeman and spin splitting of levels in magnetic field, both parallel and perpendicular to the plane of 2DES, as well as in Kondo-effect experiments. The structure of levels could be also studied by Raman spectroscopy.

The method of the paper can be used for studying systems with more particles and/or in a non-parabolic confinement potential \cite{Mikhailov01b}. It can also be used for investigating other properties of the system, for instance, the influence of impurities, or response of the dot to external fields. It is seen already now, for instance, that an asymmetrically located impurity will qualitatively differently affect the ground state of the dot at $\lambda<\lambda_c$ and at $\lambda>\lambda_c$: in the former (the latter) case the ground state is degenerate (non-degenerate) with respect to $L_{tot}$, and the impurity will result in a {\em splitting} (a {\em shift}) of the ground state level. 

To summarize, I have performed a complete theoretical study of a system of three Coulomb-interacting electrons in a parabolic confining potential, and investigated in detail physical properties and the origin of the Fermi liquid -- Wigner solid transition in the ground state of the three-electron parabolic quantum dot.

\acknowledgments
The work was supported by the Sonderforschungsbereich SFB 484, University of Augsburg. I thank Klaus Ziegler, Ulrich Eckern, Vladimir Sablikov, Teun Klapwijk and Miodrag Kulic for useful discussions, as well as referees of the paper for useful comments.

\begin{table}
\caption{Energies of the states (1,1/2) and (0,3/2) calculated in this work (exact diagonalization) and in Ref. \cite{Egger99b} (QMC, multilevel blocking algorithm).
\label{tab1}}
\begin{tabular}{rrrrr}
$\lambda$ & $S_{tot}=3/2$ & $S_{tot}=3/2$\tablenotemark[1] & $S_{tot}=1/2$ & $S_{tot}=1/2$\tablenotemark[1] \\
\tableline
2 & 8.3221 & 8.37(1)  & 8.1651 & 8.16(3)\\
4 & 11.0527 & 11.05(1) & 11.0422 & 11.05(2)\\
6 & 13.4373 & 13.43(1) & 13.4658 & no data\\
8 & 15.5938 & 15.59(1) & 15.6334 & no data\\
10& 17.5863 & 17.60(1) & 17.6279 & no data\\
\end{tabular}
\tablenotetext[1]{Ref. \cite{Egger99b}}
\end{table}


\begin{thebibliography}{10}

\bibitem{Jacak97}
L. Jacak, P. Hawrylak, and A. Wojs, {\em Quantum Dots} (Springer, Berlin,
  1997).

\bibitem{Ashoori96}
R.~C. Ashoori, Nature {\bf 379},  413  (1996).

\bibitem{Kouvenhouven97}
L.~P. Kouvenhoven {\it et~al.},  in {\em Mesoscopic Electron Transport}, edited
  by L.~L. Sohn, L.~P. Kouvenhoven, and G. Sch\"on (Kluver Academic Publisher,
  Dordrecht, 1997), Vol.~345.

\bibitem{Heitmann93}
D. Heitmann and J.~P. Kotthaus, Physics Today {\bf 46},  56  (June 1993).

\bibitem{Schuller98}
C. Sch\"uller,  in {\em Festk\"orperprobleme/Adv. in Solid State Physics},
  edited by B. Kramer (Vieweg, Braunschweig, 1998), Vol.~38, p.\ 167.

\bibitem{Taut93}
M. Taut, Phys. Rev. A {\bf 48},  3561  (1993).

\bibitem{Taut94}
M. Taut, J. Phys. A.: Math. Gen. {\bf 27},  1045  (1994).

\bibitem{Turbiner94}
A. Turbiner, Phys. Rev. A {\bf 50},  5335  (1994).

\bibitem{Dineykhan97}
M. Dineykhan and R.~G. Nazmitdinov, Phys. Rev. B {\bf 55},  13707  (1997).

\bibitem{Maksym90}
P.~A. Maksym and T. Chakraborty, Phys. Rev. Lett. {\bf 65},  108  (1990).

\bibitem{Merkt91}
U. Merkt, J. Huser, and M. Wagner, Phys. Rev. B {\bf 43},  7320  (1991).

\bibitem{Pfannkuche91}
D. Pfannkuche and R.~R. Gerhardts, Phys. Rev. B {\bf 44},  13132  (1991).

\bibitem{Wagner92}
M. Wagner, U. Merkt, and A.~V. Chaplik, Phys. Rev. B {\bf 45},  1951  (1992).

\bibitem{Hawrylak93}
P. Hawrylak and D. Pfannkuche, Phys. Rev. Lett. {\bf 70},  485  (1993).

\bibitem{Hawrylak93b}
P. Hawrylak, Phys. Rev. Lett. {\bf 71},  3347  (1993).

\bibitem{MacDonald93}
A.~H. MacDonald and M.~D. Johnson, Phys. Rev. Lett. {\bf 70},  3107  (1993).

\bibitem{Yang93}
S.-R. E. Yang, A.~H. MacDonald, and M.~D. Johnson, Phys. Rev. Lett. {\bf 71},
  3194  (1993).

\bibitem{Pfannkuche93}
D. Pfannkuche, V. Gudmundsson, and P.~A. Maksym, Phys. Rev. B {\bf 47},  2244
  (1993).

\bibitem{Palacios94}
J.~J. Palacios {\it et~al.}, Phys. Rev. B {\bf 50},  5760  (1994).

\bibitem{Peeters96}
F.~M. Peeters and V.~A. Schweigert, Phys. Rev. B {\bf 53},  1468  (1996).

\bibitem{Ezaki97}
T. Ezaki, N. Mori, and C. Hamaguchi, Phys. Rev. B {\bf 56},  6428  (1997).

\bibitem{Eto97}
M. Eto, Jpn. J. Appl. Phys. {\bf 36},  3924  (1997).

\bibitem{Creffield99}
C.~E. Creffield, W. H\"ausler, J.~H. Jefferson, and S. Sarkar, Phys. Rev. B
  {\bf 59},  10719  (1999).

\bibitem{Reimann00}
S.~M. Reimann, M. Koskinen, and M. Manninen, Phys. Rev. B {\bf 62},  8108
  (2000).

\bibitem{Yannouleas00}
C. Yannouleas and U. Landman, Phys. Rev. Lett. {\bf 85},  1726  (2000).

\bibitem{Maksym96}
P.~A. Maksym, Phys. Rev. B {\bf 53},  10871  (1996).

\bibitem{Bolton96}
F. Bolton, Phys. Rev. B {\bf 54},  4780  (1996).

\bibitem{Harju98}
A. Harju, V.~A. Sverdlov, and R.~M. Nieminen, Europhys. Lett. {\bf 41},  407
  (1998).

\bibitem{Harju99a}
A. Harju, V.~A. Sverdlov, R.~M. Nieminen, and V. Halonen, Phys. Rev. B {\bf
  59},  5622  (1999).

\bibitem{Harju99b}
A. Harju, S. Siljam\"aki, and R.~M. Nieminen, Phys. Rev. B {\bf 60},  1807
  (1999).

\bibitem{Egger99a}
R. Egger, W. H\"ausler, C.~H. Mak, and H. Grabert, Phys. Rev. Lett. {\bf 82},
  3320  (1999).

\bibitem{Egger99b}
R. Egger, W. H\"ausler, C.~H. Mak, and H. Grabert, Phys. Rev. Lett. {\bf 83},
  462(E)  (1999).

\bibitem{Pederiva00}
F. Pederiva, C.~J. Umrigar, and E. Lipparini, Phys. Rev. B {\bf 62},  8120
  (2000).

\bibitem{Filinov01}
A.~V. Filinov, M. Bonitz, and Y.~E. Lozovik, Phys. Rev. Lett. {\bf 86},  3851
  (2001).

\bibitem{Hirose99}
K. Hirose and N.~S. Wingreen, Phys. Rev. B {\bf 59},  4604  (1999).

\bibitem{Steffens98a}
O. Steffens, U. R\"ossler, and M. Suhrke, Europhys. Lett. {\bf 42},  529
  (1998).

\bibitem{Steffens98b}
O. Steffens, U. R\"ossler, and M. Suhrke, Europhys. Lett. {\bf 44},  222
  (1998).

\bibitem{Steffens98c}
O. Steffens, M. Suhrke, and U. R\"ossler, Physica B {\bf 256-258},  147
  (1998).

\bibitem{Wensauer00}
A. Wensauer, O. Steffens, M. Suhrke, and U. R\"ossler, Phys. Rev. B {\bf 62},
  2605  (2000).

\bibitem{Maksym95}
P.~A. Maksym, Europhys. Lett. {\bf 31},  405  (1995).

\bibitem{Ruan95}
W.~Y. Ruan, Y.~Y. Liu, C.~G. Bao, and Z.~Q. Zhang, Phys. Rev. B {\bf 51},  7942
   (1995).

\bibitem{Hausler96}
W. H\"ausler, Z. Phys. B {\bf 99},  551  (1996).

\bibitem{Yannouleas99}
C. Yannouleas and U. Landman, Phys. Rev. Lett. {\bf 82},  5325  (1999).

\bibitem{Hausler00}
W. H\"ausler, Europhys. Lett. {\bf 49},  231  (2000).

\bibitem{Taut00}
M. Taut, J. Phys. Condens. Matter {\bf 12},  3689  (2000).

\bibitem{Reusch01}
B. Reusch, W. H\"ausler, and H. Grabert, Phys. Rev. B {\bf 63},  113313
  (2001).

\bibitem{Maksym00}
P.~A. Maksym, H. Imamura, G.~P. Mallon, and H. Aoki, J. Phys. Condens. Matter {\bf 12},  R299 (2000).

\bibitem{Tanatar89}
B. Tanatar and D.~M. Ceperley, Phys. Rev. B {\bf 39},  5005  (1989).

\bibitem{Chui95}
S.~T. Chui and B. Tanatar, Phys. Rev. Lett. {\bf 74},  458  (1995).

\bibitem{Abrahams01}
E. Abrahams, S.~V. Kravchenko, and M.~P. Sarachik, Rev. Mod. Phys. {\bf 73},
  251  (2001).

\bibitem{Fock28}
V. Fock, Z. Phys. {\bf 47},  446  (1928).

\bibitem{Darwin31}
C.~G. Darwin, Cambridge Phil. Soc. {\bf 27},  86  (1931).

\bibitem{Bolton93}
F. Bolton and U. R\"ossler, Superlatt. Microstruct. {\bf 13},  139  (1993).

\bibitem{Bedanov94}
V.~M. Bedanov and F.~M. Peeters, Phys. Rev. B {\bf 49},  2667  (1994).

\bibitem{Mikhailov01b}
S.~A. Mikhailov, preprint {\em cond-mat/0106369} (unpublished).

\end{thebibliography}
\end{document}